# Observation of novel interference patterns in $Bi_{2-x}Fe_xTe_{3+d}$ by Fourier transform scanning tunneling spectroscopy

Y. Okada[1], C. Dhital[1], Wen-Wen Zhou[1], Hsin Lin[2], S. Basak[2], A. Bansil[2], Y. -B. Huang[3], H. Ding[3], Z. Wang[1], Stephen D. Wilson[1] & V. Madhavan[1]

In topological insulators (TI), strong spin-orbit coupling results in non-trivial scattering processes of the surface states, whose effects include suppressed back scattering[1,2,3,4] weak anti-localization[5,6] and the possibility of an exotic Kondo effect that mimics graphene[7]. Introducing time reversal breaking perturbations and establishing long-range magnetic order has been theorized to lead to the formation of quantized magnetoelectric phenomena[8], fractionalized charge excitations, and the appearance of quantum wire states[9]. A key, elusive, step in exploring these and other novel electronic phases is the experimental observation of charge backscattering due to spin-flip processes at the surface of a magnetically-doped TI. Here we utilize Fourier transform scanning tunneling spectroscopy (FT-STS) to probe the surface of a magnetically doped TI, $Bi_{2-x}Fe_xTe_{3+d}$. Our measurements show the appearance of a hitherto unobserved channel of electronic backscattering along surface q-vector ΓK. By combining FT-STS with angle-resolved photoemission data, we identify the momentum space origins of the observed q-vectors, which on comparison to a model calculation are enhanced by spin-flip scattering. Our findings present compelling evidence for the first spatial and momentum resolved measurements of magnetic impurity induced backscattering in a prototypical TI.

[1]Department of Physics, Boston College, Chestnut Hill, Massachusetts 02467, USA.

[2]Physics Department, Northeastern University, Boston, Massachusetts 02115, USA

[3]Beijing National Laboratory for Condensed Matter Physics, and Institute of Physics, Chinese Academy of Sciences, Beijing 100190, China

When spin-orbit coupling is strong enough, a band-parity inversion can be generated around the direct conduction-valence band gap of an insulator. Such topological insulators support gapless surface states that are topologically protected in the absence of time-reversal (TR) symmetry breaking. Since massless Dirac fermions are helicity eigenstates, these surface states in the low energy and long-wavelength limit have linear energy-momentum dispersions and carry a helical spin-texture: the electron's spin is normal to the direction of its lattice momentum. Direct experimental determination of the allowed scattering processes of the surface electrons is an important step in establishing their predicted spin-texture and its role within the protected surface phase. Specifically, TR symmetry prohibits backscattering of electrons carrying opposite spins. Indeed, the backscattering wave vector has not been present in the interference patterns in previous STM studies of TR invariant TIs. However, an equally important prediction is that when TR symmetry is broken, by intentionally added magnetic impurities such as Fe and Mn for example, backscattering is allowed due to the presence of spin-flip scattering. Whereas intensive STM[1, 2, 3, 4] and ARPES[10, 11, 12] studies have beautifully demonstrated many of the canonical properties of TIs, to date no direct observations of TR violating scattering vectors (q-vectors) have been reported on the surface of magnetically perturbed TIs.

In this study, we focus on Fe-doped $Bi_2Te_3$ single crystals with a nominal Fe doping of 0.25%, $(Bi_{1-x}Fe_x)_2Te_3$. $Bi_2Te_3$ belongs to a new generation of TR invariant 3D topological insulators of the type $Bi_2X_3$ (X= Te, Se etc) whose discovery has been pivotal in this field[13, 14, 15, 16]. In the topological insulator $Bi_2Te_3$, a warping of the surface state due to the three-fold symmetric crystal potential and interaction with the bulk bands provides an enhancement of certain scattering channels over others, allowing them to be observed by STM. In the absence of TR symmetry, this warping alone may lead to nesting-induced spin density wave order[17]. This renders $Bi_2Te_3$ an ideal starting point for utilizing FT-STS to resolve the influence of magnetic impurities on scattering processes within the TR invariant surface phase.

The Fe-doped $Bi_2Te_3$ single crystals were grown via a modified Bridgman technique whose details are in supplementary information. Samples were cleaved in ultra-high vacuum and immediately inserted into the cold STM head where data were obtained at 4.2 K. ARPES data were obtained at the SIS beamline of the Swiss Light Source at temperatures of 10K or 20 K, energy resolution 10-17 meV, and momentum resolution better than 0.02 $A^{-1}$.

$Bi_2Te_3$ cleaves between the quintuple layers terminating in a Te surface. As reported by other groups, various kinds of defects are observed (Fig. 1a) that may be attributed to Bi substitutional impurities in the Te plane, Te defects in the Bi plane, or vacancies in either plane. For our studies, we have the deliberately doped Fe atoms, which are expected to enter substitutionally in the Bi plane but may also appear as defects in the Te plane. FFT of the STM images show the hexagonal lattice associated with the Te atoms (inset Fig. 1a). dI/dV spectra reveal a suppression of density of states near the Fermi energy (Fig. 1b). While it is not straightforward to extract the position of the conduction or valance bands from the density of states spectra, we will later show that the shape of the spectrum is consistent with the position of the bulk bands extrapolated from prior ARPES data.

To obtain the surface state dispersion, *dI/dV(r, eV)* maps were obtained at various energies above and below the Fermi energy $E_F$. In order to maximize momentum resolution in the FFT, all data in this paper are from large area maps (a minimum linear dimension of 1500 Å). Similar to previous studies, interference patterns (IP) emerge above a threshold energy, 60 meV above $E_F$ for our samples (Figs. 2a and 2b). Unlike prior studies of $Bi_2Te_3$, neither a step edge nor a surface deposition of impurities was necessary to observe the IP. The in-plane impurities and defects sufficiently scatter the surface state electrons. Fourier transforms (FFT) of the dI/dV maps show a six-fold symmetric pattern. At energies above 150 meV we see clear intensity centered along the ΓM directions ($q_{\Gamma M}$) (inset to Fig. 2b). The appearance of a ΓM scattering vector has been previously reported in $Bi_2Te_3$ samples with no TR symmetry breaking impurities. While the helical spin-texture (inset to Fig. 1d, also see Fig. 4) prohibits direct backscattering, other scattering vectors are indeed allowed, some of which become visible in STM once hexagonal warping sets in. It is important to note that there is as yet no consensus on the momentum space (k-space) origin of this ΓM q-vector either theoretically[17, 18, 19] or experimentally[2, 4].

Remarkably, at low energies starting around 60 meV, our data reveal a new set of scattering vectors centered along ΓK ($q_{\Gamma K}$), which have not been previously observed (inset to Fig. 2a). Since the FFT shows non-dispersive features at low energies, a division by the lowest energy (60 meV) FFT makes the patterns above 60 meV clearer (Fig. 2c). Plotting the dispersion obtained from the raw undivided FFT (see supplementary Fig. S2 for evolution of raw FFT with energy) along ΓM (Fig. 2e) and ΓK (Fig. 2f) we observe that their slopes are clearly different. We now discuss the possible origin of these dispersive scattering vectors in STM data.

We first consider the q-vectors along the primary directions ΓM and ΓK. While there are a multitude of scattering channels, a few of the special wave vectors (labeled $q_1$ to $q_6$) are shown for each direction in the inset to Fig. 3a. The STM-FFTs are dominated by a subset of these vectors. $q_3$ and $q_6$ serve as upper bounds on the magnitude of the vectors in the ΓM direction and ΓK direction respectively while $q_1$ is a lower bound on the magnitude of the scattering vector along ΓM. $q_5$ along ΓK and $q_3$ along ΓM are forbidden in the time reversal invariant TI.

In the presence of TR symmetry breaking Fe impurities however, these vectors must be considered.

In order to identify the origin of $q_{\Gamma M}$ and $q_{\Gamma K}$ we compare STM dispersion to ARPES. Since the IP occur above $E_F$ in our samples, our ARPES data below $E_F$ do not capture this energy region. Fortuitously, in $Bi_2Te_3$, ARPES studies indicate that doping rigidly shifts the position of the Fermi energy with respect to the surface and bulk bands leaving the relative position of these bands unchanged.[13] We therefore access the higher energy dispersion from previously published ARPES on Sn doped $Bi_2Te_3$ where the Dirac point marked by $E_D$ could be as much as -340 meV below the Fermi level[13]. We fit the dispersion along $\Gamma M$ and $\Gamma K$ (fitting details in supplementary information) and use this to calculate the different q-vectors as a function of energy. The dispersions of the six q-vectors ($q_1$ to $q_6$) obtained by this method are plotted in figure 3a.

Overlaying ARPES q-vectors on $q_{\Gamma M}$ and $q_{\Gamma K}$ (Fig. 3a) we find that the slope, direction, and values of $q_{\Gamma M}$ and $q_{\Gamma K}$ impose rather strict constraints on their origin and allow a much more precise determination of $E_D$. The best match between ARPES and STM dispersions for $q_{\Gamma M}$ and $q_{\Gamma K}$ is obtained for $E_D$ = -105 ± 5 meV. This value matches well with ARPES data obtained by us on the same samples (Fig. 1c) and lends support to our identification of $q_{\Gamma M}$ and $q_{\Gamma K}$. With a precise knowledge of $E_D$, the relative energies of the top of the valance band and the bottom of the conduction band (Fig. 1d) can be extrapolated from ARPES. We find that our spectral shape corresponds well with the band structure, showing a minimum in the density of states close to the valance band maximum and a Van-Hove feature associated with the conduction band minimum along $\Gamma M$. It is interesting to note that since that a 3D band edge does not normally produce sharp features in the density of states; the Van-Hove feature associated with the bottom of the conduction band is due to the rather flat-band dispersion of the latter near the Γ point shown in Fig. 1d.

From the dispersion, $q_{\Gamma M}$ can be identified with scattering vectors close to $q_1$ (Fig. 3b), a scattering channel not suppressed by the spin-texture (Fig. 4c). $q_{\Gamma K}$ on the other hand originates from vectors close to $q_5$ and $q_6$ (Fig. 3c) one of which ($q_5$) would be strictly forbidden in the time reversal invariant TI. To qualitatively explain our observation of these particular q-vectors in STM, we consult the topology of the surface state band. From ARPES[13], the constant energy contours (CEC) of the surface state band change from an isotropic circular shape into a hexagon and then finally into the six-pointed shape of a snowflake (Fig. 4a, b and c). Shifting the ARPES determined energies[13] to correspond with $E_D$ = -105 meV, we find that the transition from circular to hexagon-shaped CEC would begin at approximately +50 meV while the distorted snowflake shape appears near the bottom of the conduction band (~+140 meV). The first appearance of $q_{\Gamma K}$ (~ +60 meV) therefore coincides with hexagon-shaped CECs. In this regime, there is little difference between $q_6$ and $q_5$ (Fig. 4b). The important point however is that these scattering vectors are enhanced at the stationary points of the distorted (hexagonal or snowflake shape) constant energy contours[17] which provide nesting or near nesting conditions. Since the surface state bands in this material have rather low density of states, such nesting (or near nesting) conditions are critical to observing IP in these materials. The same is true for $q_{\Gamma M}$, which appears after the CEC distorts further from a hexagon into a snowflake (>+140 meV), where it arises from the most strongly nested regions (Fig. 4c).

While the intensity of the FFT patterns is centered along the high symmetry directions, these are not the only q-vectors observed in our STM data. In fact our FFT images show extended hexagon-shaped patterns (Fig. 5a and Fig. S2), rather than point like features implying that q-vectors other than those strictly along ΓM and ΓK are involved. To determine the generators of these extended features we perform a model calculation of the expected FFT patterns both with and without spin-texture related matrix elements (all possible scattering vectors allowed) as shown in figure. 4. We fit a k.p model[17] that captures the $\int_k A(k)T(q,k)A(k+q)dk$ single Dirac-cone surface states centered at the Gamma point to ARPES data[13] (Fig. S3). This allows us to obtain numerical values for the parameters in the model (details in supplementary information). To calculate the FFT patterns, we use the spin-dependent scattering probability SSP(q) = where $A(k)$ is the spectral weight and $T(q,k)$ is a scattering matrix element.

We find that there are strong systematic similarities between the STM and the calculated FFT patterns. First, the extended hexagon shape of the STM FFT-patterns (Figs. 2c, d and Fig. 5a) compare well with calculations (Fig 4f-i) and arise from the fact that small angle deviations from the primary scattering vectors along $q_{\Gamma M}$ and $q_{\Gamma K}$ are still strongly nested. Importantly, upon comparing the low-energy FFT in the energy regime of the hexagonal CEC, we find that the STM data (Fig. 2c) are most consistent with the theoretical FFT with no spin-texture related matrix elements (Fig. 4h). In fact, in the absence of TR symmetry breaking, matrix element effects almost completely suppress $q_6$ and the nearby vectors (Fig. 4f) observed in STM. Our data therefore clearly indicate that Fe impurities facilitate new backscattering channels, which are otherwise suppressed in samples with TR symmetry. A close examination of the STM q-vectors (fig. 3b and c) show that at energies below 140 meV the direct backscattering spin-flip channel $q_5$ is a better fit to STM data, however above 140 meV, our data are consistent with the average of $q_5$ and $q_6$. It is remarkable that the subtle differences between $q_5$ and $q_6$ can be distinguished in STM data.

We now discuss our main observations and their implications. First, we find two dominant scattering channels in the STM data, which arise from stationary points on the constant energy contours in k-space. Our analysis shows that the new ΓK channel at low energies can be associated with spin-flip backscattering vectors close to $q_5$, which are enhanced by the Fe-impurities. Second, consistent with our calculations, the STM FFT patterns are extended hexagons rather than point-like indicating that scattering vectors within an angular range of the high symmetry directions are observed. Third, comparing the STM data to ARPES allows us to identify not only the Dirac point but also the Van-Hove feature associated with the conduction band minimum close at the Γ point. Tracking the energy of this feature in STM spectra therefore provides a convenient method to identify positions of the bulk bands and Dirac point with respect to the Fermi energy for a given $Bi_2Te_3$ sample.

Given the constraints on the surface state dispersion from ARPES and the exceedingly good agreement between STM and ARPES, the identification of $q_{\Gamma M}$ with the nested set of vectors $q_1$ is inescapable. This raises intriguing questions. As seen in ARPES data, the tips of the snowflake become increasingly ill-defined at higher energies as the surface state enters the bulk band. Remarkably, we observe a linear dispersion for this set of vectors up to 600 meV above the Dirac point well into the bulk conduction band (Fig. 5b). This seems to suggest that the surface states remain distinct (possibly as a resonance) over a surprisingly large energy range such that the bulk-surface mixing continues to be suppressed beyond the insulating gap. We expect that continued investigations via FT-STS within the higher energy regime where the $q_{\Gamma M}$ dispersion deviates from linearity would provide important information on the high-energy electronic structure in topological insulators and serve as a vital complement to ARPES measurements currently precluded from probing this range.

Figure 1

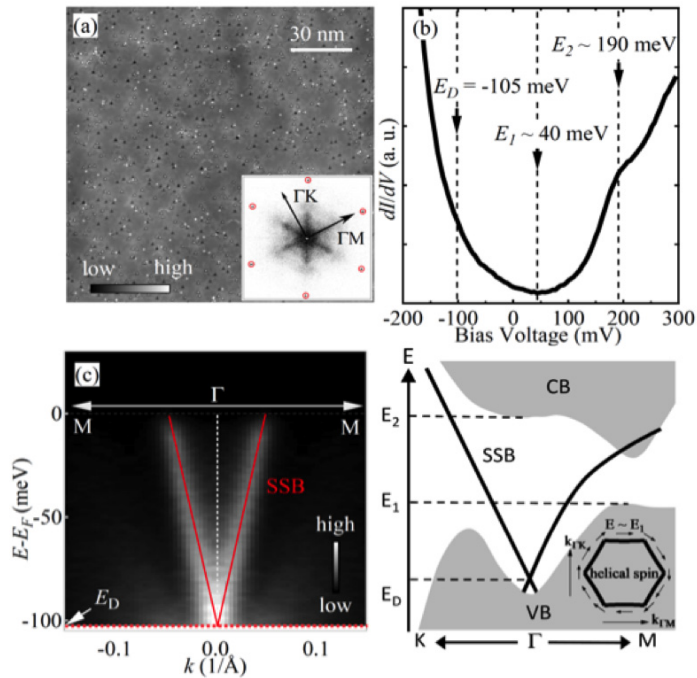

**Fig. 1 STM and ARPES spectroscopy of Fe-doped Bi$_2$Te$_3$.** a) STM topography (linear dimension 1500 Å) showing the various triangle shaped impurities and defects. The inset is the FFT of the image, which clearly shows the hexagonal pattern of the atoms. The corresponding high-symmetry directions in k-space are specified. For the remainder of the paper all k-space figures are rotated such that ΓM is along the horizontal axis. b) dI/dV spectrum showing the energies $E_D$ $E_1$ and $E_2$ schematically represented in 1d. $E_D$ is from the STM data. The other two points were extrapolated from prior ARPES. Upon extrapolation, we notice that $E_2$ is clearly visible in STM data. c) ARPES on samples from the same batch showing the DP close to -100 meV. d) Schematic of band structure and surface state showing the energies of the Dirac point ($E_D$), the valance band top along ΓM ($E_1$) and conduction band bottom at the Γ point. Since the Dirac point varies slightly between samples, all data in this paper is from different maps on one particular sample. The other Fe-doped samples measured by us showed similar results. All energies in this paper are with reference to the Fermi energy of the sample ($E_F$ = 0 meV).

Figure 2

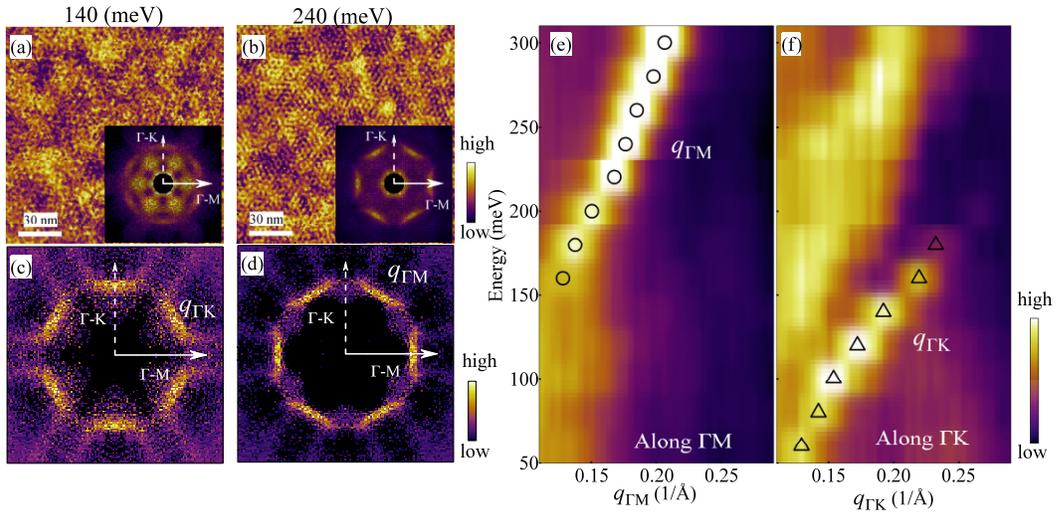

**Fig. 2 dI/dV maps and Fourier transforms, q-space data from STM.** a) and b) *dI/dV (r, E)* maps showing the IP with the inset showing the raw (undivided) FFT. c) and d) Same FFT as inset but divided by FFT at +60 meV. All FFT patterns in this paper are hexagonally symmetrized. e) and f) Intensity profiles of FFTs (linecuts of FFTs from the Γ point in the center in a particular direction, represented as a color intensity). These are plotted at various energies (vertical axis) along ΓM (e) and ΓK (f) and show the evolution with energy of the q-vectors $q_{\Gamma M}$ (circle) and $q_{\Gamma K}$ (triangle) defined in (c) and (d).

Figure 3

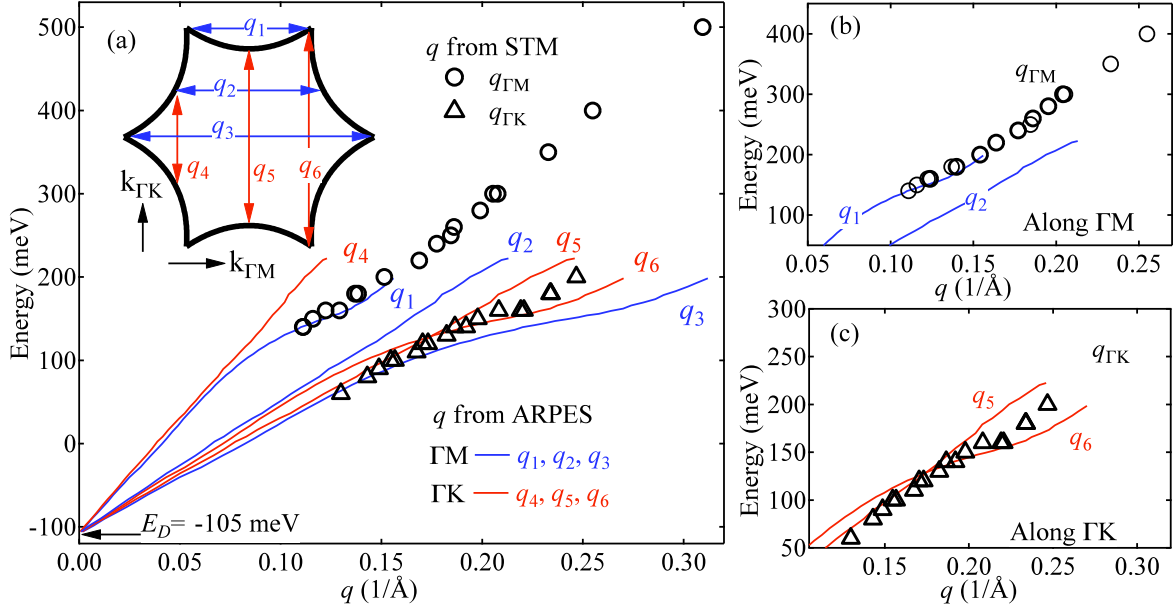

**Fig. 3 Comparison of STM and ARPES q-space dispersion.** a) Dispersion of the six q-vectors shown in the inset calculated from ARPES data[13] (solid lines) and shifted in energy to match the STM q-vectors obtained from the FFT of the IP (circles and triangles). Data is shown from multiple dI/dV maps on the same sample. The best match between STM and ARPES is obtained for a Dirac point of -105 meV. b) and c) Zoom in of the dispersion of $q_{\Gamma M}$ and $q_{\Gamma K}$. The tight constraints imposed by the direction, magnitude and slope of the q-vectors results in the identification of $q_{\Gamma M}$ with vectors close to $q_1$, and $q_{\Gamma K}$ with $q_5$ or $q_6$ (which are almost identical at low energies in the hexagonal CEC regime). Note that the crossover from hexagon-like to snowflake-like CECs occurs at the energy where $q_5$ and $q_6$ cross in 3c (~140 meV), while the change from circular to hexagonal CECs occurs above the energy where $q_6$ starts to deviate from linearity in 3a (~50 meV).

Figure 4

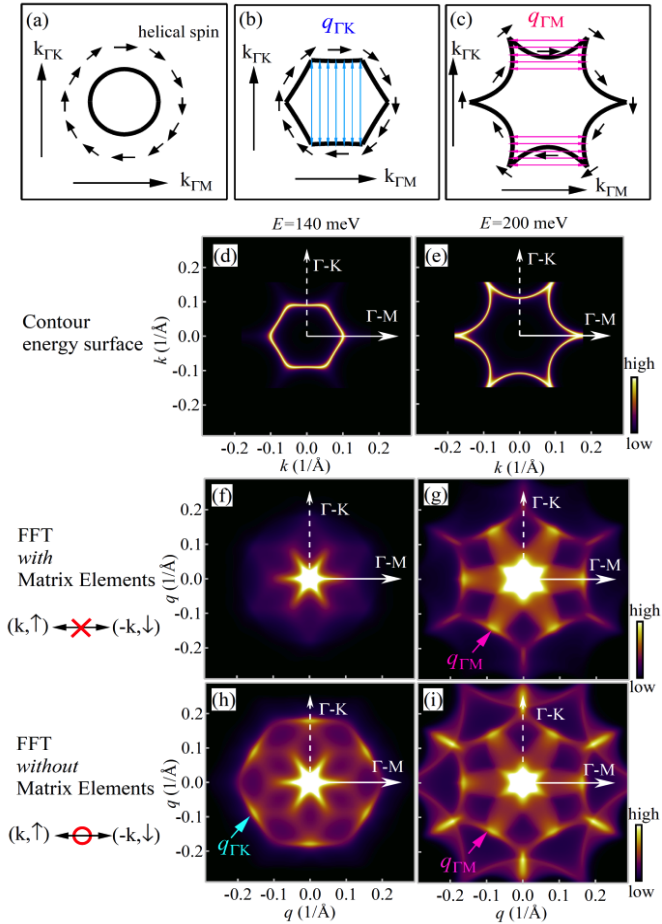

**Fig. 4 Calculated FFT patterns with and without spin-texture.** a), b), and c) Schematic of CECs and the nested regions of the CECs resulting in the primary q-vectors observed in STM at low energies along ΓK and higher energies along ΓM. d) and e) Calculated constant energy contours at two representative energies, obtained from combining ARPES data along ΓM and ΓK with a theoretical form for the k-dependence of the CECs. f) and g) The calculated FFT obtained from (d) and (e) including spin texture related matrix elements. h) and i) The calculated FFT obtained from (d) and (e) with no spin texture related matrix elements (spin flip allowed).

Figure 5

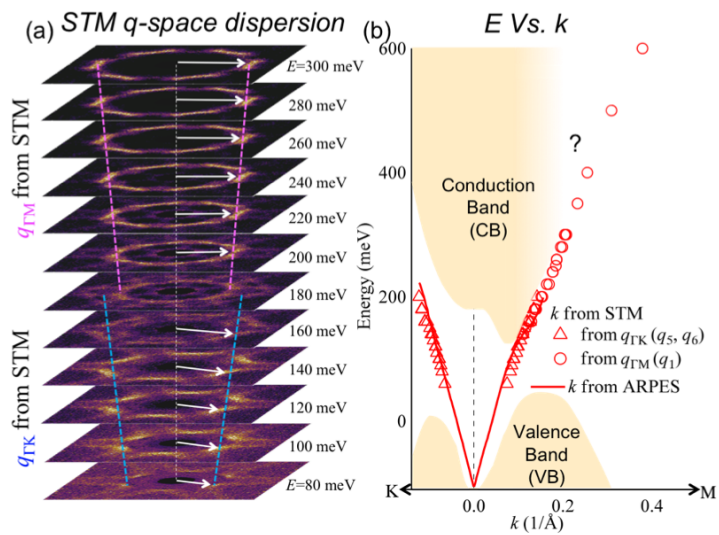

**Fig. 5 k-space dispersion from STM compared with ARPES.** a) Evolution of STM q-vectors with energy. At each energy, the FFT pattern obtained by STM was divided by the FFT at +60 meV. b) Identifying the origin of the STM q-vectors allows us to translate back into k-space. From $q_{\Gamma M}$ we obtain k along $\Gamma M$ (red circles). From Fig. 3 we see that $q_{\Gamma K}$ may be associated with either $q_5$ or $q_6$. Associating $q_{\Gamma K}$ with $q_5$ we can calculate k along $\Gamma K$ while associating it with $q_6$ allows us to plot k along $\Gamma M$ (red triangles in both directions). Since $q_5$ or $q_6$ have a similar magnitude at low energies the error in calculating k is small. Yet, we notice that the overall fit to ARPES data is better along $\Gamma K$ indicating that the dominant channel at low energies is the spin-flip $q_5$. The red solid line is extrapolated from ARPES data and the band structure is a schematic based on ARPES.

# References

[1] Roushan, P. *et al*. Topological surface states protected from backscattering by chiral spin texture. *Nature*. **460**, 1106-1109 (2009).

[2] Zhang, T. Experimental demonstration of topological surface states protected by time-reversal symmetry. *Phys. Rev. Lett.* **103**, 266803 (2009).

[3] Gomes, K. K. *et al*. Quantum imaging of topologically unpaired spin-polarized Dirac fermions. Preprint at <http://arxiv.org/abs/0909.0921v2> (2009).

[4] Alpichshev, Z. *et al*. STM imaging of electronic waves on the surface of $Bi_2Te_3$: topologically protected surface states and hexagonal warping effects. *Phys. Rev. Lett.* **104**, 016401 (2010).

[5] Chen, J. *et al*. Gate-voltage control of chemical potential and weak antilocalization in $Bi_2Se_3$. *Phys. Rev. Lett.* **105**, 176602 (2010).

[6] He, H. T. *et al*. Impurity effect on weak anti-localization in topological insulator $Bi_2Te_3$. Preprint at <http://arxiv.org/abs/1008.0141v1> (2010).

[7] Tran, M. T. & Kim, K. S. Probing surface states of topological insulator: Kondo effect and Friedel oscillation under magnetic field. Preprint at http://arxiv:1006.3208v2 (2010).

[8] Qi, X. L., Hughes, T. & Zhang, S. C. Topological field theory of time-reversal invariant insulators. *Phys. Rev. B* **78**,195424 (2008).

[9] Nomura, K. & Nagaosa, N. Electric charging of magnetic textures on the surface of a topological insulator. *Phys. Rev. B* **82**, 161401 (2010).

[10] Hsieh, D. *et al*. A tunable topological insulator in the spin helical Dirac transport regime. *Nature* **460**, 1101-1105 (2009).

[11] 11. Zhang, H. J. *et al*. Topological insulators in $Bi_2Se_3$, $Bi_2Te_3$ and $Sb_2Te_3$ with a single Dirac cone on the surface. *Nature Phys*. **5**, 438-442 (2009).

[12] Hsieh, D. *et al*. Observation of unconventional quantum spin textures in topological insulators. *Science* **323**, 919-922 (2009).

[13] Chen, Y. L. et al. Experimental realization of a three-dimensional topological insulator, $Bi_2Te_3$. *Science* **325**, 178 (2009).

[14] Fu, L., Kane, C. L. & Mele, E. J. Topological insulators in three dimensions. Phys. Rev. Lett. 98, 106803 (2007).

[15] Moore, J. E. & Balents, L. Topological invariants of time-reversal-invariant band structures. *Phys. Rev. B* **75**, 121306(R) (2007).

[16] Hsieh, D. *et al*. Observation of time-reversal-protected single-dirac-cone topological-insulator states in $Bi_2Te_3$ and $Sb_2Te_3$. *Phys. Rev. Lett*. **103**, 146401 (2009).

[17] Fu, L. Hexagonal warping effects in the surface states of the topological insulator $Bi_2Te_3$. *Phys. Rev. Lett*. **103**, 266801 (2009).

[18] Lee, W. C., Wu, C., Arovas, D. P. & Zhang S. C. Quasiparticle interference on the surface of the topological insulator $Bi_2Te_3$. *Phys. Rev. B* **80**, 245439 (2009).

[19] Zhou, X., Fang, C., Tsai, W. F. & Hu, J. Quasiparticle scattering in two dimensional helical liquid. Preprint at <http://arxiv:0910.0756v3> (2009).